\documentclass[10pt]{article}
\usepackage{amssymb}

{\hspace*{\fill}$\rule{.3\baselineskip}{.35\baselineskip}$\end{trivlist}}

\makeatletter
\@addtoreset{equation}{section}
\makeatother

\newcommand{\Z}{\mathbb{Z}}

\newfam\bifam
\font\tenbi=cmmib10 scaled \magstep1
\font\sevenbi=cmmib10 at 11pt
\font\fivebi=cmmib10 at 6pt
\textfont\bifam = \tenbi
\scriptfont\bifam= \sevenbi
\scriptscriptfont\bifam= \fivebi

\usepackage[dvips]{epsfig}
\usepackage{graphicx}

\begin{document}

\title{{\bf Statics and Dynamics of an Inhomogeneously-Nonlinear Lattice}}
\author{Debra L. Machacek$^{1}$, Elizabeth A. Foreman$^{1}$, 
Q.E. Hoq$^2$, \\
P.G. Kevrekidis$^1$, A. Saxena$^3$, D.J. Frantzeskakis$^4$ 
and A.R. Bishop$^3$ \\
{\small $^{1}$ Department of Mathematics,
University of Massachusetts, Amherst, Massachusetts, 01003-4515, USA} \\
{\small $^{2}$ Department of Mathematics,
Western New England College, Springfield, MA, 011119, USA} \\
{\small $^{3}$ Theoretical Division and Center for Nonlinear Studies,
Los Alamos National Laboratory, Los Alamos, NM 87545, USA} \\
{\small $^4$ Department of Physics, University of Athens, Panepistimiopolis,
Zografos, Athens 15784, Greece} \\
}
\date{\today}
\maketitle

\begin{abstract}
We introduce an inhomogeneously-nonlinear Schr{\"o}dinger lattice,
featuring a defocusing segment, a focusing segment and a
transitional interface between the two. We illustrate that
such inhomogeneous settings present vastly different dynamical behavior
than the one expected in their homogeneous counterparts in
the vicinity of the interface. We analyze the relevant stationary
states, as well as their stability by means of perturbation theory
and linear stability analysis. We find good agreement with the
numerical findings in the vicinity of the anti-continuum limit.
For larger values of the coupling, we follow the relevant branches
numerically and show that they terminate at values of the coupling
strength which are larger for more extended solutions. The dynamical 
development of relevant instabilities is also monitored in the case 
of unstable solutions.
\end{abstract}

\section{Introduction}

In the past two decades, the number of applications of discrete,
nonlinear dynamical models has increased dramatically.
A diverse set of applications has emerged, 
ranging from the nonlinear
optics of guided waves in inhomogeneous optical
structures \cite{EMSAPA02,PMAAESPL02} and photonic crystal lattices
\cite{ESCFS02,SKEA03}, to atomic physics and the dynamics of
Bose-Einstein condensate (BEC) droplets in periodic (optical lattice)
potentials \cite{CBFMMTSI01,CFFFMI03,ABDKS01,AKKS02} and from
condensed matter, in Josephson-junction ladders \cite{F03,MO03}, to
biophysics, in various models of double-stranded DNA 
\cite{DPB93,PDHW93}. This broad span of research areas and
corresponding applications has now been summarized in a variety
of reviews 
\cite{A97,FW98,HT99,KRB01,EJ03}.

A model that has drawn a particular focus among these areas
of applications is the so-called discrete nonlinear Schr{\"o}dinger
equation (DNLS) \cite{KRB01}. This model was first proposed in
the context of nonlinear optics, where it describes beam dynamics in coupled waveguide arrays \cite{CJ88,CLS03},
but is equally applicable in other settings, such as the dynamics
of BECs confined 
in deep optical lattices \cite{AKKS02,TS01}. 
The relevant model involves the nearest
neighbor coupling between adjacent waveguides (wells of the optical lattice in BECs)
and the local nonlinear self-action
induced by the Kerr effect in each waveguide (or the mean-field inter-atomic interaction in the condensate setting).

Typically, the above setup is homogeneous in that all waveguides
or wells are identical. However, recently there has been a surge of 
activity motivated by the experimental tunability of the properties of 
individual waveguides/wells. In particular, in the optical setting,
the interaction of discrete solitary waves with structural
defects was examined in \cite{ol03}, while ``non-standard'' solitary
waves (discrete gap solitons) were observed in binary waveguide
arrays \cite{ol04,prl04}. This activity has been recently reviewed
in \cite{opto05} discussing various aspects of ``optics in 
non-homogeneous waveguide arrays''. On the BEC side, there are also
similar developments involving not only the (attractive or
repulsive) localized ``defect'' action of a laser beam on the
condensate \cite{kett1,kett2}, but also the potential of creating the so-called ``superlattice''
structures by means of the superposition of optical potentials 
of different periodicity \cite{phillips}.

In this context of 
inhomogeneous nonlinear dynamical systems, we propose here a novel setting, which
we illustrate to have drastically different dynamical behavior
than that we would expect from its homogeneous counterparts. 
In particular, we impose a spatial pattern on the nonlinearity,
having the form of an ``interface'' between a set of defocusing Kerr 
waveguides on the one end and a set of focusing Kerr waveguides
on the other, separated by a ``transient'' layer (interface) of
a waveguide bearing intermediate properties between the two 
segments above. This is, in some aspects, reminiscent 
to the recent proposition in the context of BECs of spatially dependent nonlinearities
(see e.g. \cite{smel1,smel2,bor1} and references therein). We show
that this setting already presents a wealth of static and dynamical behavior 
which is very different than its homogeneous analog.

We focus on the localized, solitary wave excitations 
in the vicinity of the interface. Starting from the so-called
anti-continuum limit of zero coupling \cite{MA94}, we show that
the existence and stability of the localized solutions in the vicinity
of the interface can be quantified for low couplings
by means of a perturbation theory using the coupling constant 
as small parameter. 
As the coupling between
the sites near the interface increases, the phenomenology becomes
drastically different, leading the relevant solution branches
to a termination through saddle-node bifurcations that would
be absent in the corresponding homogeneous limit. Perhaps equally
surprisingly, the more extended multi-pulse solutions appear to 
survive for larger values of the coupling than the single (or smaller
size) pulse waves, which is again contrary to what is expected
from the homogeneous limit. In this stronger coupling regime,
we investigate the properties of the relevant solutions through
numerical bifurcation theory and linear stability analysis. We use 
direct numerical simulations to illustrate the manifestations
of the dynamical instabilities of those among the solutions which are
found to be dynamically unstable. We believe that this example
illustrates the rich diversity of behavior that can be 
manifested in such inhomogeneous settings.

Our presentation is structured as follows: in section 2,
we present the setup and analytical results, while in section 3, 
we study the model numerically and compare with the analytical
results. Finally, in section 4, we summarize our findings and
present our conclusions.

\section{Setup and Analytical Results}

We consider an inhomogeneous lattice model 
described by a discrete non-linear Schr{\"o}dinger equation of the following form,
\begin{equation}
i \frac{d u_n}{dt}=- C \Delta _2u_n-d_n|u_n|^2u_n,
\label{1}
\end{equation}
where $C$ is the coupling between the adjacent sites of the lattice, 
$\Delta_2 u_n = (u_{n+1}+u_{n-1}-2 u_n)$ is the discrete Laplacian.
The evolution
variable is $z$ in the optical case and $t$ in the BEC case; for notational
simplicity, we use $t$ in what follows.
The nonlinearity coefficient \(d_n\) (where \(n\) is the spatial index) is determined by 
the intensity-dependent refractive index (in the context of optics) 
or the s-wave scattering length (in the context of BECs) 
of each waveguide (or optical lattice well for BECs).
The center site, \(d_0\), 
is assumed to have an intermediate value slightly greater than zero. 
For all $n<0$, the refractive index is set to the
defocusing value \(d_n=-0.9\). For all $n>0$, the refractive index is 
set to the focusing value \(d_n=1.1\). Note that $d_0$ is set to the average
of these two values i.e., to $0.1$; it should also be mentioned that the results reported
below were found to be typical of similar choices of the $d_n$ profile.

We focus our attention on standing wave
solutions of the form of \(u_n=e^{i \Lambda t}v_n\), where $\Lambda$ is the propagation constant 
in optics or the chemical potential in BECs, and 
\(v_n\) is the spatial (time-independent) profile, satisfying the steady state equation: 
\begin{equation}
G(v_n,C) \equiv \Lambda v_n-C \Delta _2v_n-d_n|v_n|^2v_n=0.
\label{3}
\end{equation} 
It can be easily seen (see e.g. Refs. \cite{ABK04,DKF05})
that, without loss of generality, we can restrict ourselves to the
class of real solutions of Eq. (\ref{3}). 
In the anti-continuum (AC) limit, i.e., for $C=0$, the solutions are immediately
obtainable in the form:
$v^{2}_{n}=\{0,\frac{\Lambda}{d_n}\}$, provided 
that $d_n>0$ for all $n$. Using this solution
and setting to non-zero values only specific individual 
sites to the right of the zeroth site, 
we prescribe the configurations (in the AC limit)
for the various branches that will be subsequently examined
analytically as well as numerically.
The selected configurations are as follows; 
lower 
first branch \(|1\!>\): single excited site at $n=0$; 
upper first branch \(|1,e\!\!>\): excited sites at $n=0$ and $n=1$
in phase; lower second 
branch \(|1,-e\!\!>\): excited sites $n=0$ and $n=1$ but out of phase;
 upper second branch \(|1,-e,-e\!\!>\): excited sites $n=0$ with 
positive sign and $n=1,2$ with negative signs. Following the same
pattern, the remaining branches are:
lower third branch \(|1,-e,e\!\!>\); 
upper 
third branch \(|1,-e,e,e\!\!>\); lower fourth branch \(|1,-e,e,-e\!\!>\); 
upper 
fourth branch \(|1,-e,e,-e,-e\!\!>\); lower fifth branch 
\(|1,-e,e,-e,e\!\!>\); upper 
fifth branch \(|1,-e,e,-e,e,e\!\!>\).  

On the numerical side, we analyse these branches using the
pseudo-arclength continuation method \cite{doedel}.   
This allows us to trace the  branches past fold points.
In particular, given a solution $(\vec{v}_0, C_0)$ of the equation (\ref{3})
$G(\vec{v},c)=0$ 
and a direction vector $(\dot{\vec{v}_0},\dot{C_0})$, one can determine
$(\vec{v}_1,C_1)$ by solving the following system of equations:
\begin{equation}
\begin{array}{c}
G(\vec{v}_1,C_1)=0,\\
(\vec{v}_1-\vec{v}_0) \dot{\vec{v}_0}+(C_1-C_0)\dot{C}_0-\Delta s=0,
\end{array}
\label{5}
\end{equation}
where $\Delta s$ is a (small) arclength parameter.
We use Newton's method to solve the system in Eq.\ (\ref{5}) for $(\vec{v_1},C_1)$:
\begin{equation}
\left( \begin{array}{cc}
\frac{\partial}{\partial \vec{v}} G_1 & \frac{\partial}{\partial \vec{C}} G_1 
\\ 
\dot{\vec{v}_0} & \dot{C_0} 
\end{array}
\right)
\left(
\begin{array}{c}
\vec{v}_1 - \vec{v}_0 \\
C_1 - C_0
\end{array}
\right) =- 
\left(
\begin{array}{c}
G(\vec{v}_1,C_1)\\
(\vec{v}_1-\vec{v}_0) \dot{\vec{v}_0}+(C_1-C_0)\dot{C_0}-\Delta s  
\end{array}
\right).
\label{6}
\end{equation}
The next (normalized) direction vector, $(\dot{\vec{v}_1},\dot{C_1})$, 
can be computed by solving:
\begin{equation}
\left(
\begin{array}{cc}
\frac{\partial}{\partial \vec{v}} G_1 & \frac{\partial}{\partial \vec{C}} G_1 
\\ 
\dot{\vec{v}_0} & \dot{C_0} 
\end{array}
\right)
\left(
\begin{array}{c}
\dot{\vec{v}_1} \\ \dot{C_1}
\end{array}
\right) =
\left(
\begin{array}{c}
0 \\ 1
\end{array}
\right).
\label{7}
\end{equation}

To examine the 
linear stability of the stationary solutions obtained as described above,
we use the perturbation ansatz
\begin{equation}
u_n=e^{i\Lambda t}(v_n+\epsilon a_n e^{-i\omega t}+b_n e^{i\omega ^{\ast} t}),
\label{8}
\end{equation}
where $\epsilon$ is a formal small parameter. 
By substituting Eq.\ (\ref{8}) into Eq.\ (\ref{1}) and dropping higher order
terms, the following 
system of linear stability equations is obtained:
\begin{equation}
\begin{array}{c}
\omega a_n = -c \Delta_2 a_n + \Lambda a_n -2 d_n|v_n|^2a_n- d_n v_n^2 b_n
^{\ast} ,\\
\omega ^{\ast} b_n = c\Delta_2 b_n - \Lambda b_n +2 d_n
|v_n|^2b_n +  d_n v_n^2 a_n^{\ast}.
\end{array}
\label{9}
\end{equation}
The numerical solution of the ensuing matrix eigenvalue problem
for the eigenfrequencies $\omega$ and eigenvectors $\{a_n,b_n^{\ast}\}$
can be then used to characterize the linear stability (more precisely
the spectral stability) of the solutions. Since the eigenvalues 
(eigenfrequencies) of the underlying Hamiltonian system appear in 
quartets, to ensure a spectral instability it suffices for the
above linear system to possess an eigenfrequency with a non-zero
imaginary part. When the solutions are found to be unstable, 
we use a fourth-order, direct integration scheme to examine the
dynamical evolution of the instability.

Having presented the main framework and numerical methods, we 
now turn to some analytical results. 
Our analysis will be based on perturbation theory from the 
anti-continuum limit, using the coupling strength $C$ as the
small parameter. In particular, we expand the solution as:
\begin{eqnarray}
v_n=v_n^{(0)}+ C v_n^{(1)} + O(C^2). 
\label{ceq1}
\end{eqnarray}

It is easy to check that the stability problem of Eq. (\ref{9})
can be rewritten  for the eigenvalues $\lambda=i \omega$
in the Hamiltonian form
\begin{equation}
\label{eigenvalue} {\cal J} {\cal H} \mbox{\boldmath $\psi$} =
\lambda \mbox{\boldmath $\psi$},
\end{equation}
where $\mbox{\boldmath $\psi$}$ is the infinite-dimensional
eigenvector, consisting of 2-blocks of $(u_n,w_n)^T$ (the superscript $T$ denotes transpose),  
where $a_n = u_n + i w_n$, 
$b_n = u_n - i w_n$ for the eigenvector equations (\ref{9}), ${\cal J}$
is the infinite-dimensional skew-symmetric matrix, which consists of
$2$-by-$2$ blocks of 
$$
{\cal J}_{n,m} = \left( \begin{array}{cc} 0 & 1 \\ -1 & 0
\end{array} \right) \delta_{n,m},
$$
and ${\cal H}$ is the infinite-dimensional symmetric matrix, which
consists of $2$-by-$2$ blocks of
$$
{\cal H}_{n,m} = \left( \begin{array}{cc} ({\cal L}_+)_{n,m} & 0 \\
0 & ({\cal L}_-)_{n,m} \end{array} \right).
$$
The matrices $({\cal L}_+)_{n,m}$ and $({\cal L}_-)_{n,m}$
are, in turn, defined as:
$$
\left( {\cal L}_+ \right)_{n,n} = 1 - 3 d_n v_n^2, \qquad \left(
{\cal L}_- \right)_{n,n} = 1 - d_n v_n^2, \qquad \left( {\cal
L}_{\pm} \right)_{n,n+1} = \left( {\cal L}_{\pm} \right)_{n+1,n} = -C.
$$
Similarly to the solution itself, the matrix ${\cal H}$ in
the neighborhood of the AC limit, can be expanded as
\begin{equation}
\label{matrix1} {\cal H} = {\cal H}^{(0)} + \sum_{k = 1}^{\infty}
C^k {\cal H}^{(k)},
\end{equation}
where ${\cal H}^{(0)}$ is diagonal with two blocks:
\begin{equation}
\label{energy-0}
{\cal H}_{n,n}^{(0)} = \left( \begin{array}{cc} -2 & 0 \\
0 & 0 \end{array} \right), \;\; n \in S, \qquad
{\cal H}_{n,n}^{(0)} = \left( \begin{array}{cc} 1 & 0 \\
0 & 1 \end{array} \right), \;\; n \in \Z \backslash S,
\end{equation}
where $S$ denotes the set of excited sites.
Notice that 
in the $C=0$ limit,
each excited site corresponds to a pair of zero eigenvalues
in equation (\ref{eigenvalue}), while each zero-site corresponds
to a pair of eigenvalues at $\pm 1$.

Choosing for simplicity of exposition (and without loss of
generality) $\Lambda+2C=1$ in Eq. (\ref{3}), the solution of
the leading perturbation problem in (\ref{ceq1}) is governed by the following equation: 
\begin{eqnarray}
[1-3 d_n (v_n^{(0)})^2] v_n^{(1)} = v_{n+1}^{(0)} + v_{n-1}^{(0)}. 
\label{ceq2}
\end{eqnarray}
One can apply this, e.g., for the 2-site solutions such as
$|1,e>$ and $|1,-e>$ , to obtain the leading order corrections:
\begin{eqnarray}
v_0^{(1)}=-\frac{1}{2} (\pm \sqrt{\frac{1}{d_1}} ), 
\label{ceq3}
\\
v_1^{(1)}=-\frac{1}{2} (\pm \sqrt{\frac{1}{d_0}} ), 
\label{ceq4}
\end{eqnarray}
where the sign inside the parenthesis corresponds to the 
sign of excitation of the site indexed inside the square root.
Similarly, for 3 excited sites the expressions become
\begin{eqnarray}
v_0^{(1)} &=& -\frac{1}{2} (\pm \sqrt{\frac{1}{d_1}} ), 
\label{ceq3a}
\\
v_1^{(1)} &=& -\frac{1}{2} (\pm \sqrt{\frac{1}{d_0}} \pm \sqrt{\frac{1} 
{d_2}}), 
\label{ceq4a}
\\
v_2^{(1)} &=& -\frac{1}{2} (\pm \sqrt{\frac{1}{d_1}} ).
\label{ceq5a}
\end{eqnarray}
One can correspondingly generalize these expressions for an arbitrary 
number of excited sites.

We now turn to the perturbed stability problem. The small perturbation
of size $C$ cannot render the eigenvalues of order $O(1)$ unstable. Instead,
the potentially ``dangerous'' eigenvalues for instability purposes
are those which are located at the origin of the spectral plane in
the AC limit (corresponding to the excited sites, as discussed above).
The perturbed form ${\cal H}_1$ of the matrix relevant to the stability
problem can be easily seen (from the perturbative expansion) to assume
the form
\begin{eqnarray}
{\cal H}_{n,n}^{(1)} = -2 d_n \phi_n^{(0)} \phi_n^{(1)} \left(
\begin{array}{cc} 3 & 0 \\ 0 & 1 \end{array} \right), \qquad
\label{energy1} {\cal H}_{n,n+1}^{(1)} = {\cal
H}_{n+1,n}^{(1)} = -  \left( \begin{array}{cc} 1 & 0 \\
0 & 1 \end{array} \right),
\end{eqnarray}
while all other blocks of ${\cal H}_{n,m}^{(1)}$ are zero. 
If we consider the (linearly independent) eigenvectors 
corresponding to zero eigenvalues of ${\cal H}_0$, 
${\bf f}_n$, then it was proved in \cite{DKF05} that in
order to obtain the leading correction to the (zero) eigenvalues
of the original problem, it is sufficient to consider the
reduced problem
\begin{equation}
\label{reducedeigenvalue1} {\cal M}_1 {\bf c} = \gamma_1 {\bf c},
\end{equation}
where 
\begin{equation}
\label{energyM1} \left( {\cal M}_1 \right)_{m,n} = \left( {\bf f}_m,
{\cal H}^{(1)} {\bf f}_n \right)
\end{equation}
is an $N \times N$ matrix, with $N$ being the number of excited sites.
Once the eigenvalues $\gamma_1$ of this reduced problem are obtained,
then the perturbed eigenvalues of the original problem are given
by the form $\lambda= \sqrt{C} \lambda_1 + O(C)$, where
$\lambda_1=\sqrt{2 \gamma_1}$.

One can then directly compute the matrix ${\cal M}$, e.g. in the
2-site and 3-site cases, as well as more generally, to have the forms
$$
\left( {\cal M} \right)_{n,n} = - 2 d_n v_n^{(0)} v_n^{(1)}, 
\qquad 
\left( {\cal
M} \right)_{n,n-1} = -\cos(\theta_{n-1}-\theta_n), \qquad
\left( {\cal M} \right)_{n,n+1} = -
\cos(\theta_{n+1}-\theta_n).
$$
One can then obtain the following predictions for 
the leading order eigenvalues of some of the branches discussed
above (we only explicitly present these predictions in the 2-
and 3-site cases)
\begin{eqnarray}
\lambda &=& \pm 2.69003 \sqrt{C}, 
\label{ceq5}
\\
\lambda &=& \pm 2.69003  i \sqrt{C}, 
\label{ceq6}
\end{eqnarray}
for the (unstable) $|1,e>$ and (stable, at least for small $C$) 
$|1,-e>$ modes, respectively. In the 3-site,
we have:
\begin{itemize}
\item For the branch $|1,e,e>$, two real eigenvalue pairs:
\begin{eqnarray}
\lambda_1=\pm 1.277714 \sqrt{C}, \qquad \lambda_2=\pm 3.098986 \sqrt{C};
\label{ceq7}
\end{eqnarray}
\item For the branch $|1,-e,e>$, the same eigenvalue pairs as above
but multiplied by $i$ (hence the branch is marginally stable for small $C$);
\item For the branch $|1,e,-e>$, one real and one imaginary pair
in the form:
\begin{eqnarray}
\lambda_1=\pm 1.63075 i \sqrt{C}, \qquad \lambda_2=\pm 2.428092 \sqrt{C};
\label{ceq8}
\end{eqnarray}
\item Similarly, the branch $|1,-e,-e>$ has the same eigenvalues
as $|1,e,-e>$ but multiplied by $i$ (so it is also always linearly unstable).
\end{itemize}
Notice that one can, in principle, expand this type of 
analysis to any other configuration of interest.

We now turn to numerical results in order to examine the validity of
these theoretical predictions.

\section{Numerical Results}

\begin{figure}[htb]
\centering
\includegraphics[scale=.7]{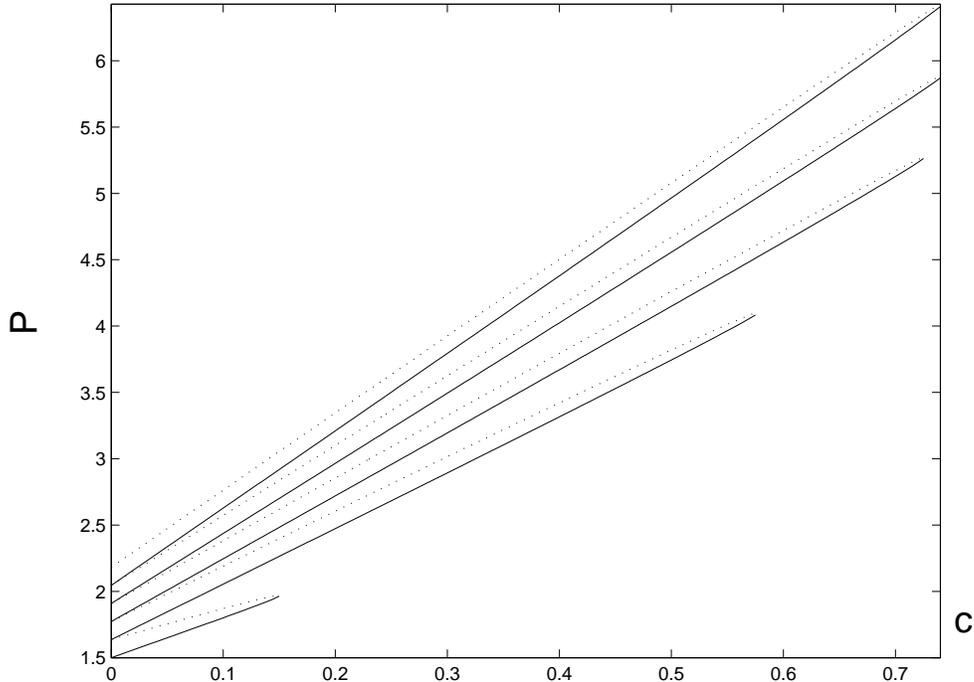}
\caption{
Bifurcation diagram of the first five branches. Plot of the solution's 
power vs. 
the continuation parameter, C. Dotted lines represent 
unstable regions. Solid lines represent initially stable regions. 
}
\label{Fig. 1}
\end{figure}

Figure \ref{Fig. 1} summarizes our essential numerical results,
presenting the squared $l^2$ norm of the solution (physically, 
the power in optics or the
rescaled number of atoms in BEC) $P=\sum_n |u_n|^2$ for the
various branches that we examined in our computations.
[For the explanation of the branches that are shown, the
reader is referred to section 2]. There is a number of features
in this bifurcation diagram which are in extreme contrast with
the corresponding homogeneous limit of this system. Firstly, the
single pulse branch in the vicinity of the interface already terminates
for quite small values of $C$; in fact, it is the first
branch to terminate in a saddle-node bifurcation with the 
two-site mode $|1,e>$. This is the analog of what would be termed
``the Page mode'' in the setting of intrinsic localized modes (ILMs). 
As the coupling increases, the site
with index $n=1$ starts becoming excited for the single pulse
branch eventually colliding (in configuration space) with
the 2-site mode and annihilating each other. In the homogeneous
limit of a focusing medium both of these branches would survive
for {\it any} C, up to the continuum limit of $C \rightarrow \infty$.
A similar phenomenology emerges for the so-called twisted
mode branch of $|1,-e>$ that, in turn, is also linearly stable
for small $C$; for larger $C$ it eventually collides with the
branch $|1,-e,-e>$ and disappears in a saddle-node bifurcation.
The same is also true for the pair of $|1,-e,e>$ and $|1,-e,e,e>$ 
and for that of $|1,-e,e,-e>$ and $|1,-e,e,-e,-e>$ also shown in
the figure. Another interesting general trend illustrated in this
diagram is that the more extended the branch (i.e., the more
sites participating in the nonlinear wave), the larger the coupling
strength for which it persists. This is also contrary to what one would
expect from the homogeneous limit, where multi-site solutions can
only be continued to a finite coupling which is typically larger for
more localized structures.  

\begin{figure}[htb]
\begin{center}
\begin{tabular}{ll}
(a)
\includegraphics[scale = .375]{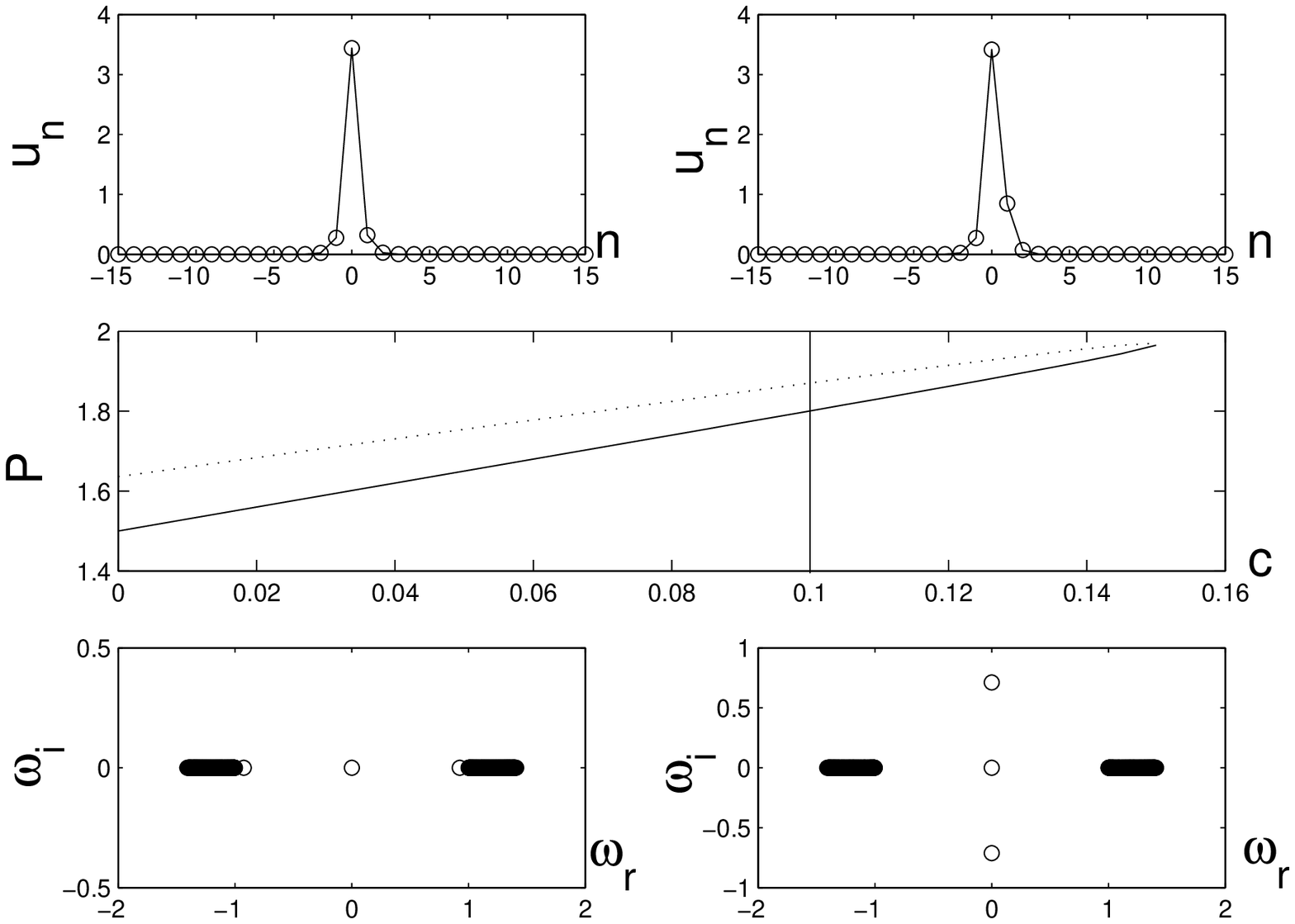}
(b)
\includegraphics[scale = .375]{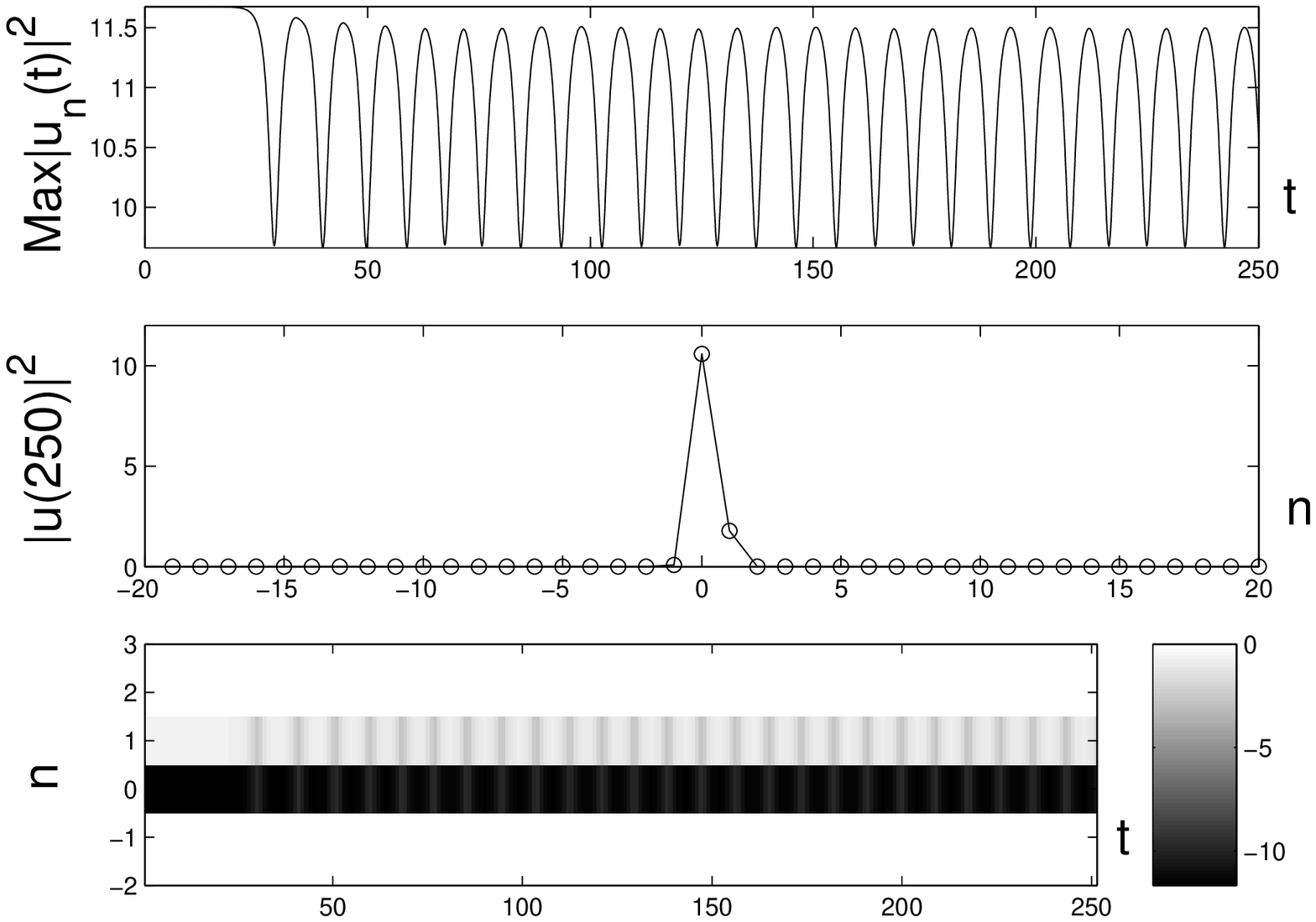}
\end{tabular}
\end{center}
\centerline{\includegraphics[scale=.45]{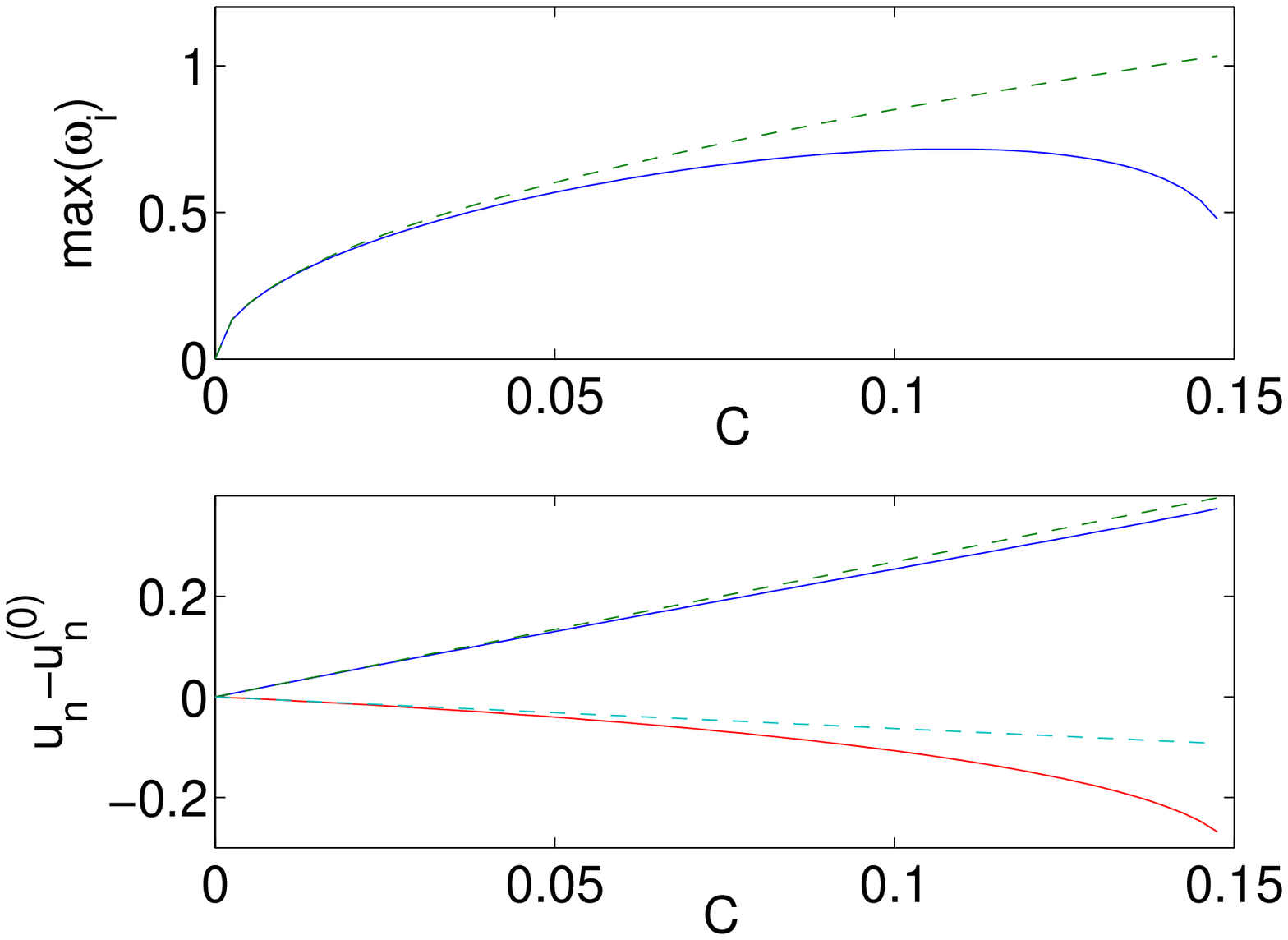}}
(c)
\caption{(a) Profiles of wave configurations and eigenfrequencies 
 at the value of $C=0.1$ where the solid vertical line crosses branch one in 
the center diagram. Upper left: wave configuration from lower half of branch 
one, $|1\!>$. Lower left: eigenfrequencies of the linearization around
this solution. 
Upper  
right: Wave configuration from upper half of branch one, $|1,e\!>$. 
Lower right:
 eigenfrequencies from the corresponding linearization. 
(b) Top: evolution of max square modulus of the solution
taken from upper branch at $C = 0.1$. Middle: spatial profile of the
square modulus 
of u taken for $c = 0.1$, after propagation by 250 units. 
Bottom: space time contour plot of the square modulus of the solution.
(c) The top panel shows the most unstable eigenvalue of the two-site
solution of $|1,e>$ as a function of the coupling strength $C$. The solid
line is the full numerical result, while the dashed line denotes the 
analytical prediction. The bottom panel shows the correction for the
central site $n=0$ and its neighboring site $n=1$ given by the first
order theory (dashed line) versus the corresponding numerical
result (solid line).}
\label{Fig. 2}
\end{figure}

Figure \ref{Fig. 2} illustrates the details of the lower
pair of branches in Fig. \ref{Fig. 1}. In particular, the 
top left panel shows the profile of the modes and their
corresponding stability for the stable $|1>$ and unstable
$|1,e>$ solutions. The continuation of the branches
(up to $C=0.15$ where they collide and disappear)
is shown in detail. The instability of the unstable
two-site solution is investigated in the right panel through
a direct simulation showing its breathing evolution. Panel (c)
reports the result of the full numerical simulation (solid line)
versus the theoretical prediction (dashed line), both for the
instability eigenvalue of $|1,e>$, 
and the profile correction imposed by Eqs. (\ref{ceq3})-(\ref{ceq4}).
It is readily observed that, for small values of $C$, the agreement between the analytical predictions and the 
numerical results is very good.
Of course, for larger $C$, the analytical results are expected to be less successful 
due to the significance of higher-order corrections neglected in our analysis.

\begin{figure}[htb]
\begin{center}
\begin{tabular}{ll}
~~~(a)~~~
{\includegraphics[scale=.35]{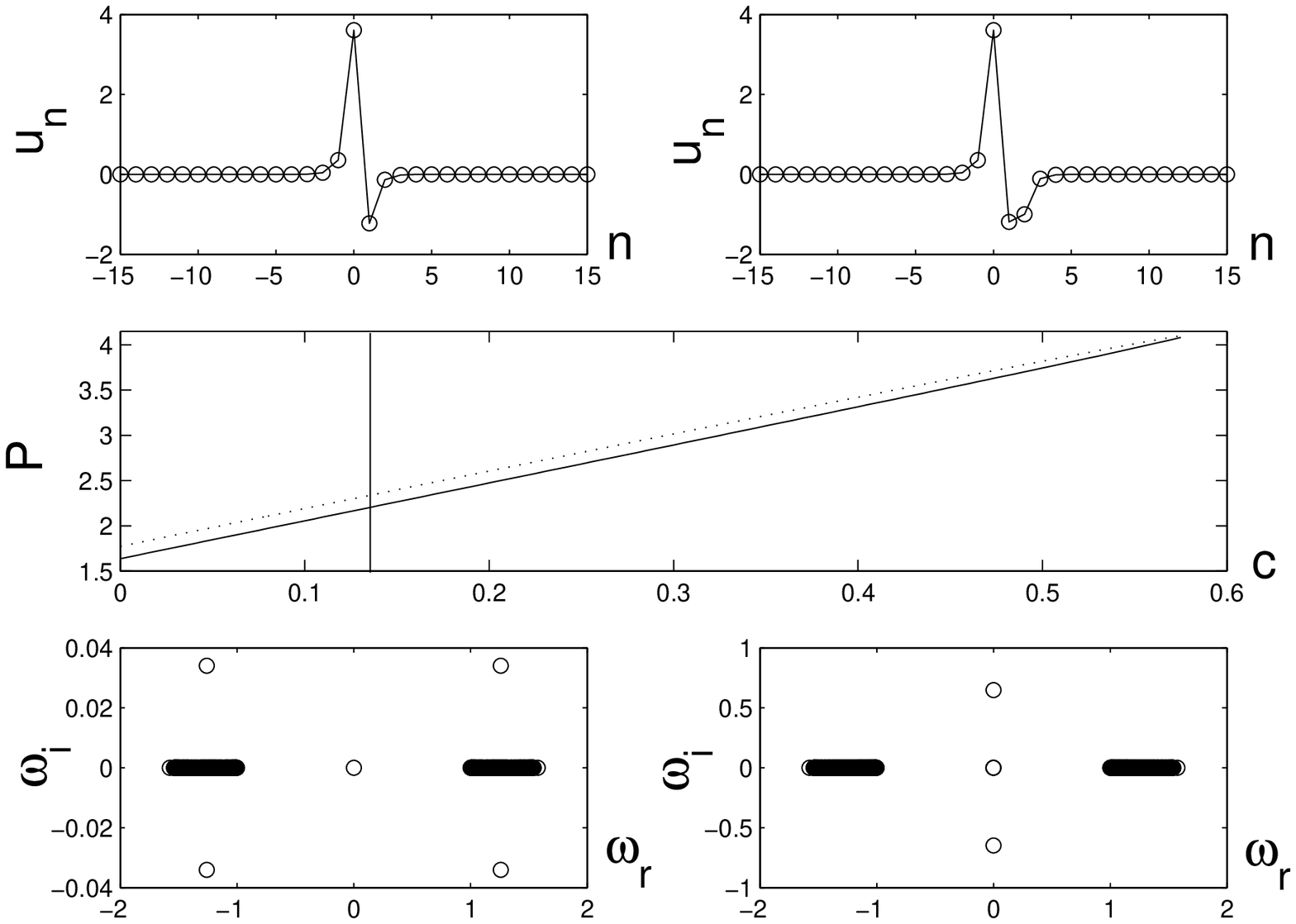}}
(b)
\includegraphics[scale=.35]{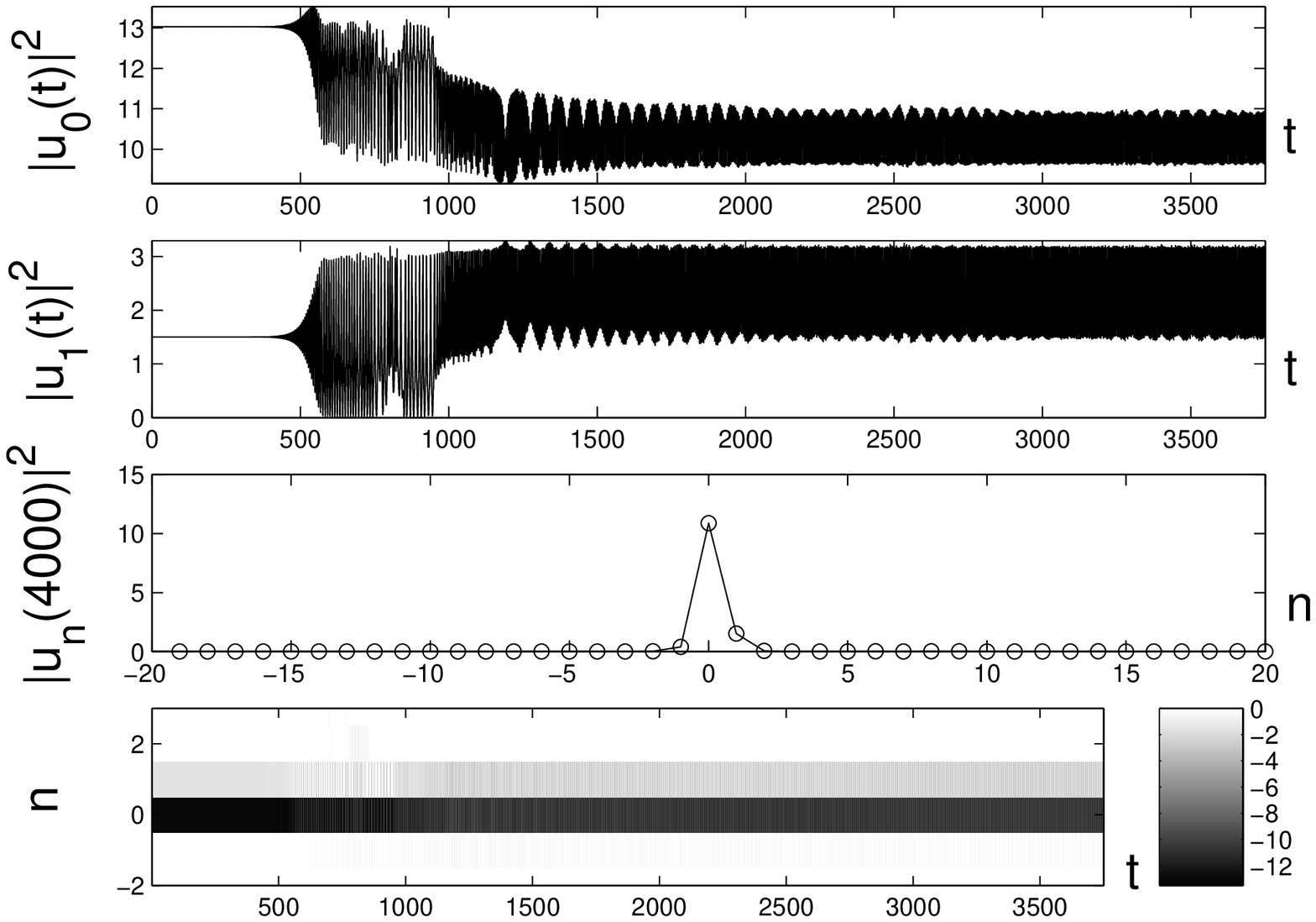}
\\
(c)
\includegraphics[scale=.35]{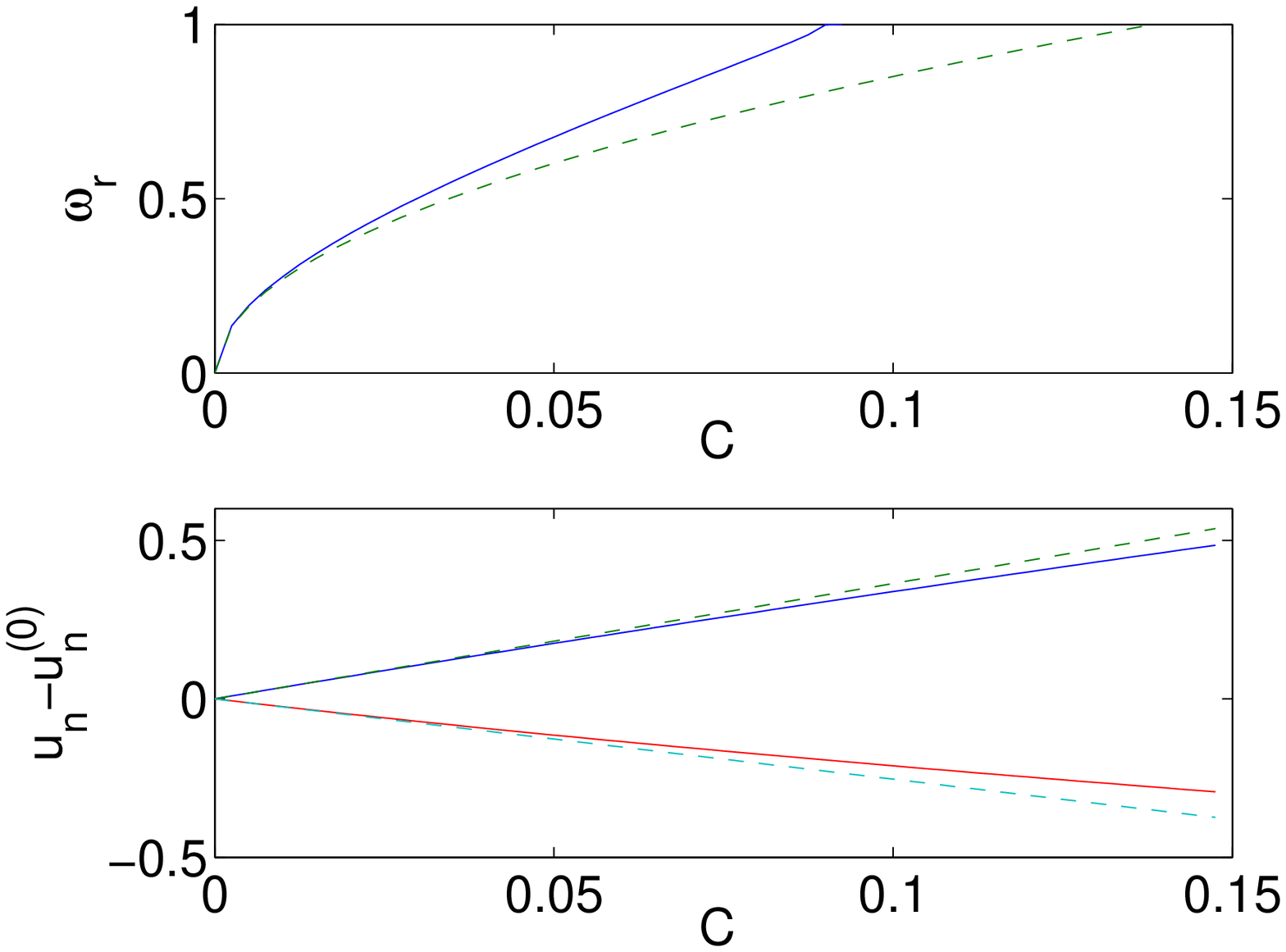}
(d) 
\includegraphics[scale=.35]{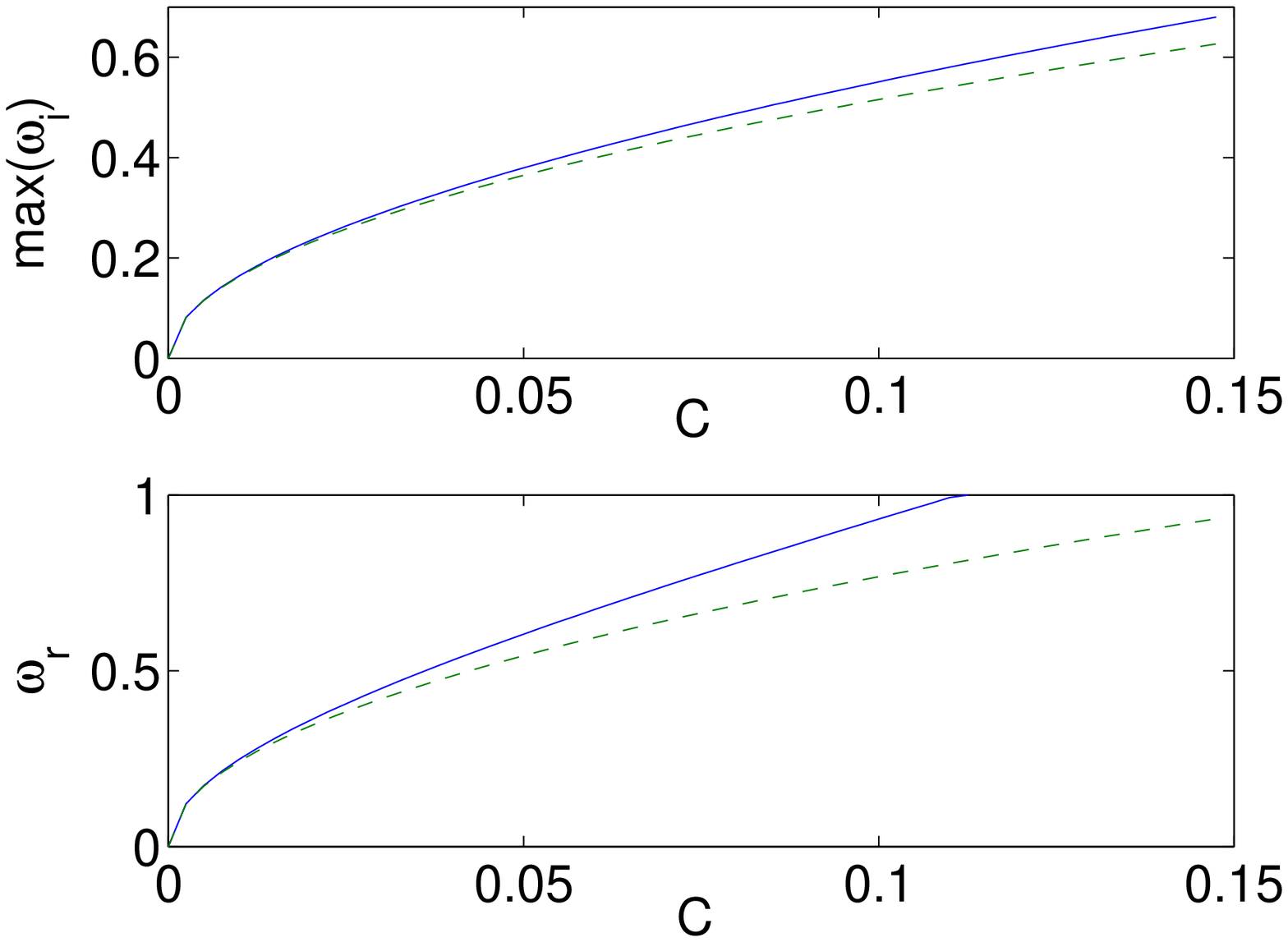}
\end{tabular}
\end{center}
\caption{(a) Similar to Figure 2 for branch two with profiles
at $C=0.135$, for the branches $|1,-e>$ and $|1,-e,-e>$.
(a)-(c) are similar as above, while panel (d) shows the eigenfrequencies
(one real and one imaginary, as predicted by theory) of
the mode $|1,-e,-e>$ from the numerical results (solid) against the
analytical predictions of section 2 (dashed lines).
 }
\label{Fig. 3}
\end{figure}

Figure \ref{Fig. 3} is similar to Fig. \ref{Fig. 2}, but for
the second pair of solutions in Fig. \ref{Fig. 1}, namely for
$|1,-e>$ and its corresponding unstable companion $|1,-e,-e>$.
These branches disappear together in a saddle-node bifurcation
for $C=0.575$. While $|1,-e,-e>$ is always unstable due to 
a real eigenvalue for any $C$, the twisted mode is stable for
small $C$, but becomes unstable for $C>0.095$ due to the collision
of two imaginary eigenvalue pairs with opposite Krein signature, 
leading to a quartet of eigenvalues through a Hamiltonian Hopf 
bifurcation \cite{vdm}. The top left panel of the figure illustrates
the solution profiles, the bifurcation diagram, and the results of the stability
analysis. 
The top right highlights the breathing
evolution of the twisted mode state. The two bottom panels 
compare the eigenvalue prediction of Eq. (\ref{ceq6}) with the
full numerical result (dashed vs. solid lines) and the 
leading order correction for the two side mode, in the case
of the left panel. In the right panel, the eigenvalue predictions
(one real and one imaginary) for the $|1,-e,-e>$ branch (dashed line) are 
also compared to the corresponding numerical results (solid line), obtaining
once again good agreement.

\begin{figure}[htb]
\begin{center}
\begin{tabular}{ll}
~~~(a)~~~
{\includegraphics[scale=.35]{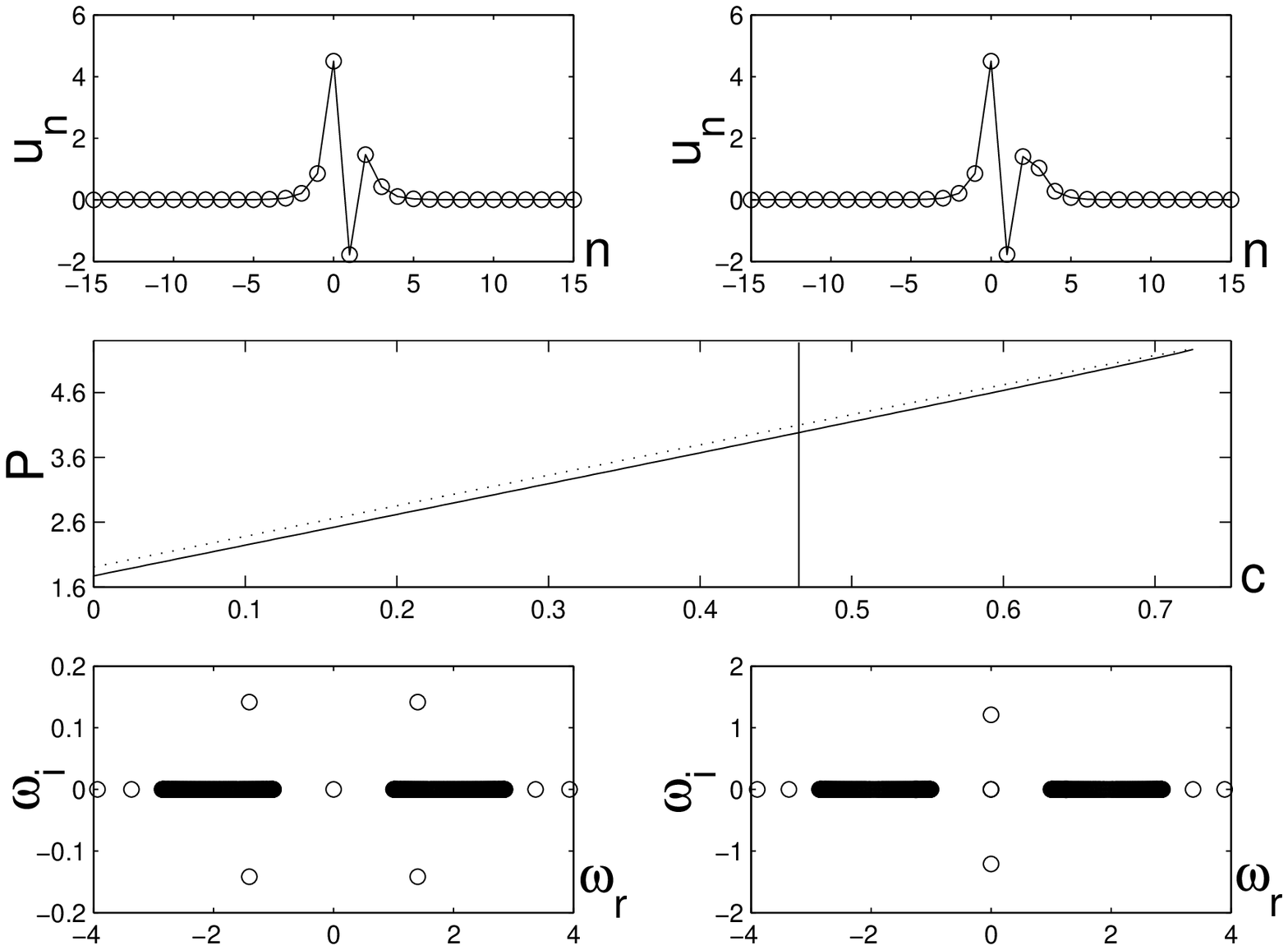}}
(b)
{\includegraphics[scale=.35]{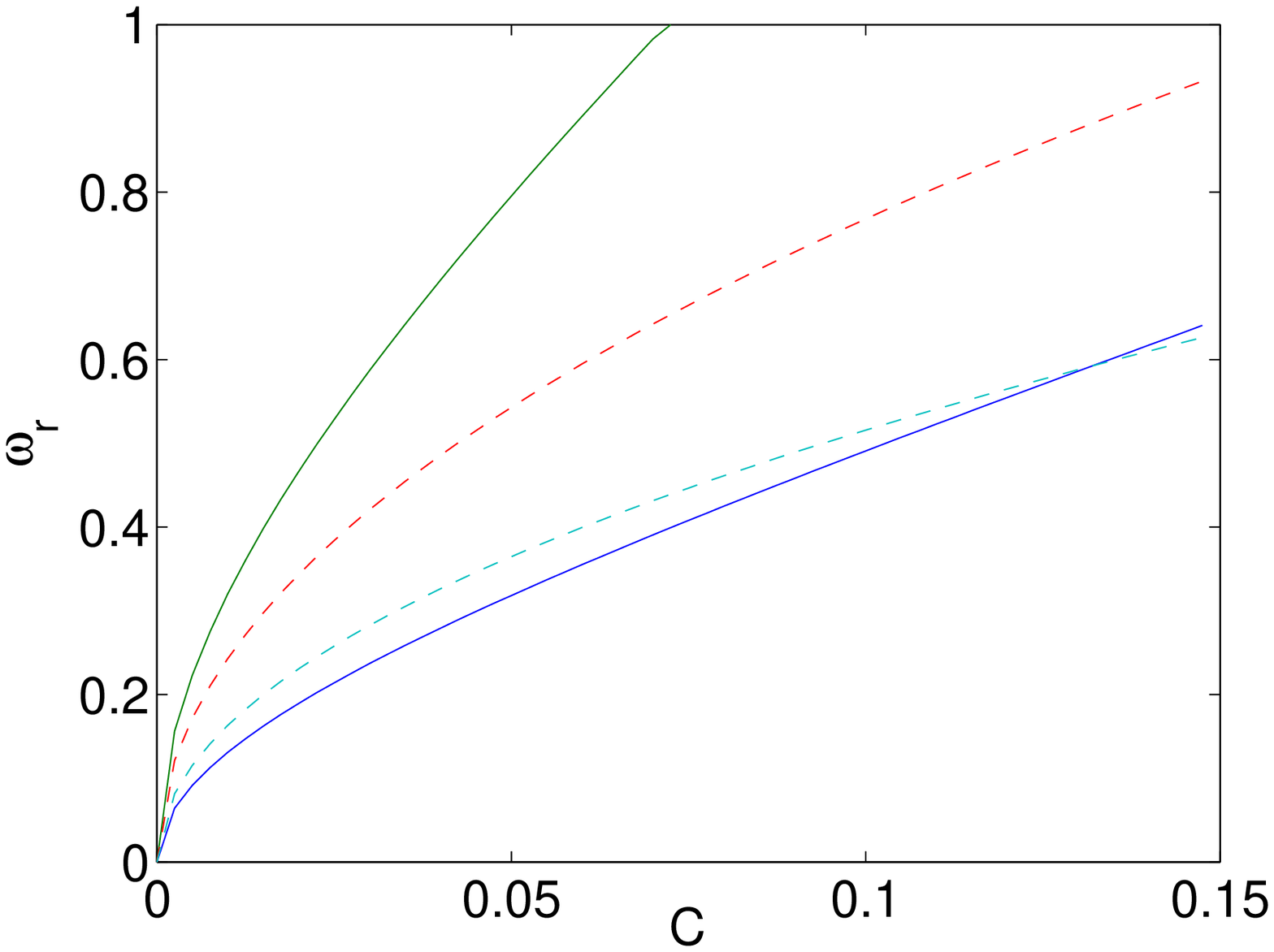}}
\\
(c)
\includegraphics[scale=.35]{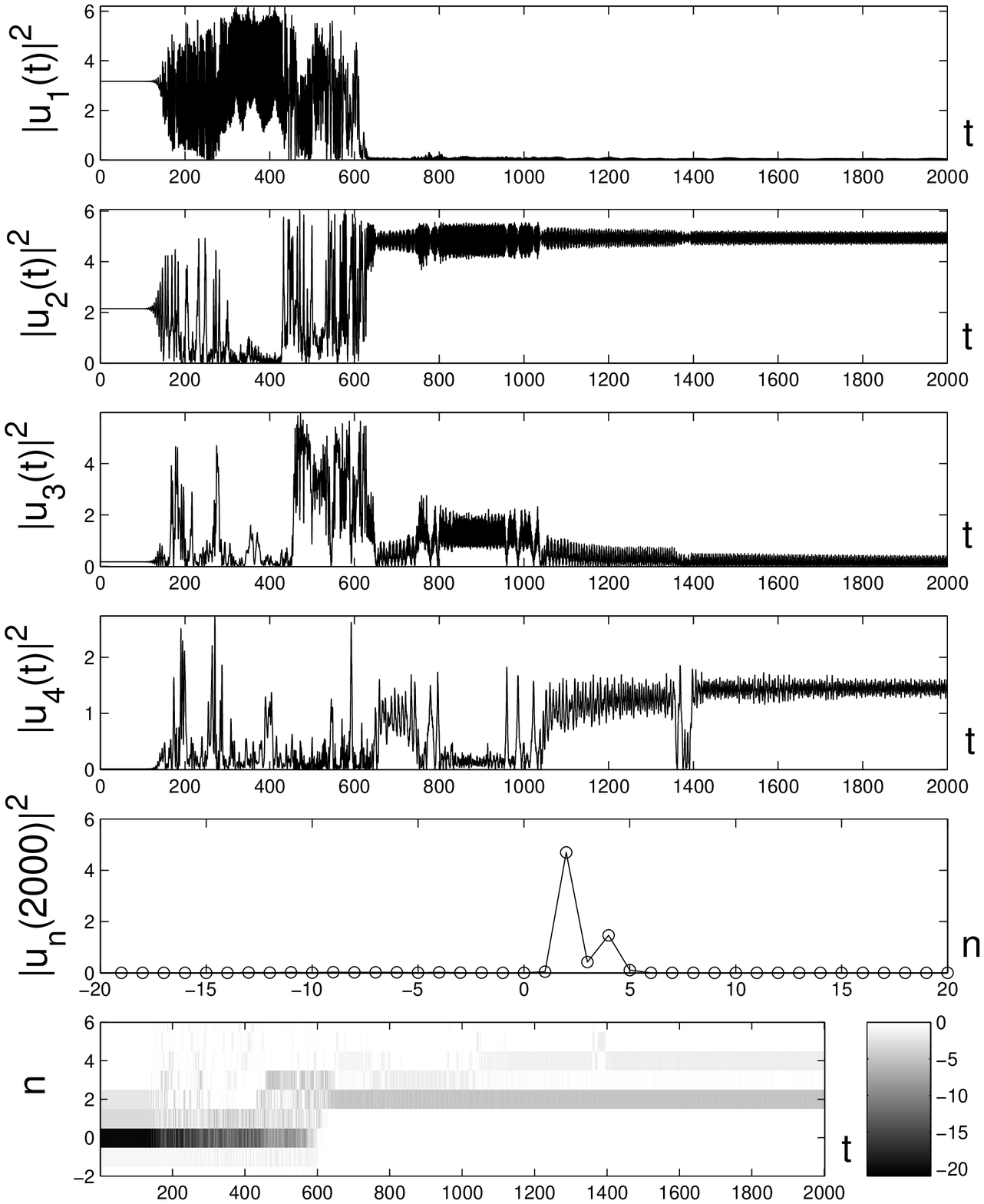}
(d)
\includegraphics[scale=.35]{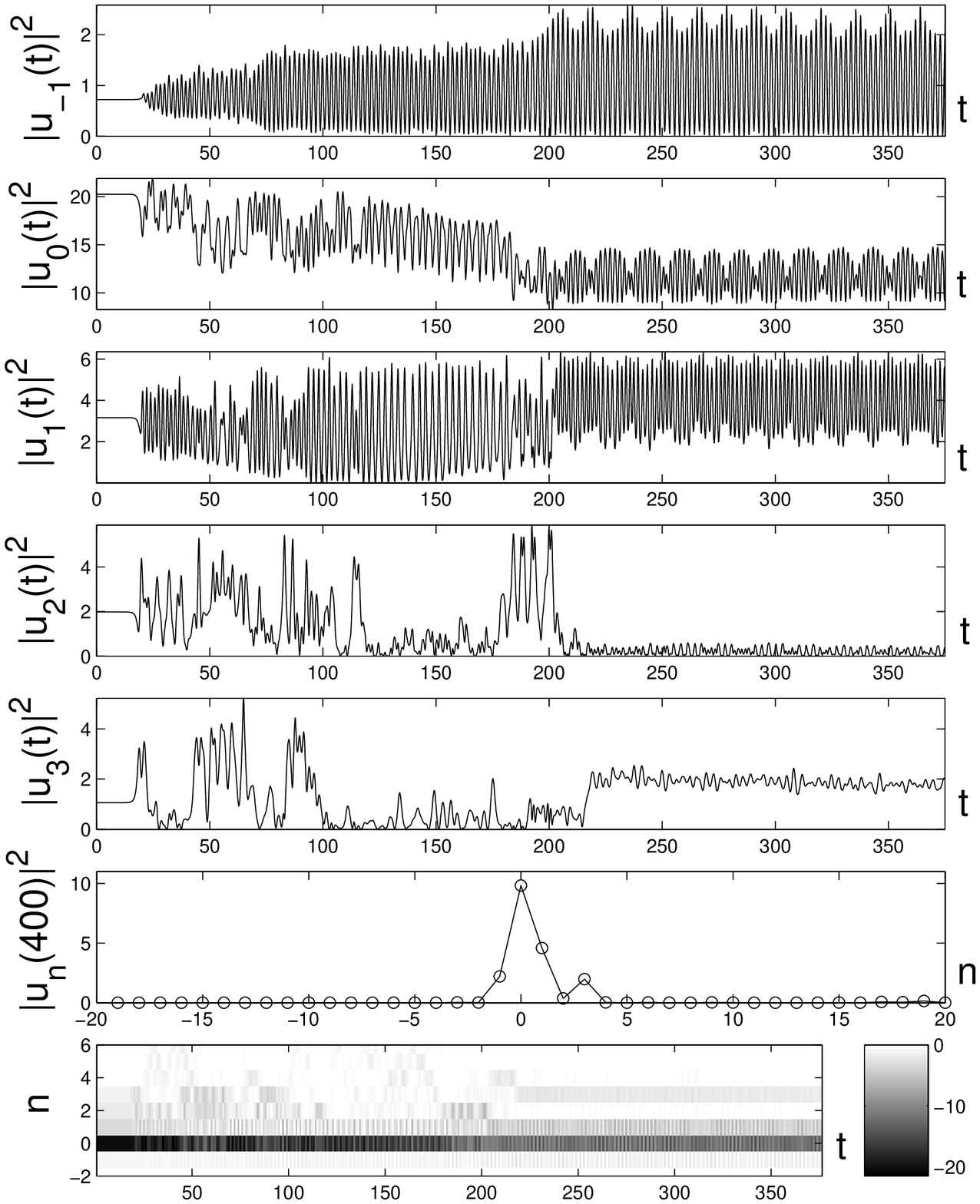}
\end{tabular}
\end{center}
\newpage
\caption{Panel 
(a) shows the profiles and eigenfrequencies of the third branch 
($|1,-e,e>$ on the left and $|1,-e,e,e>$ on the right)
for  $C=0.465$. The right panel shows, for the 3 site mode
$|1,-e,e>$, the dependence of its  two eigenfrequencies against the
analytical predictions (dashed lines). The bottom panels show,
for each of the branches and for $C=0.465$, the dynamical evolution
of the principal sites, as well as the space-time contour plot of the
square modulus.
}
\label{Fig. 4}
\end{figure}

In Figure \ref{Fig. 4}, we examine the third pair of branches of Fig.
\ref{Fig. 1},
namely (the stable for small $C$) $|1,-e,e>$ and (the always
unstable) $|1,-e,e,e>$. These branches, in turn, collide
and disappear through a saddle-node for $C=0.725$. The top right
panel shows the prediction (dashed line) versus
the numerical results (solid line) for the leading order eigenvalues of the
$|1,-e,e>$ solution (discussed in the previous section). The theory 
correctly captures, at small $C$, the existence of two imaginary eigenvalues
and their $C^{1/2}$ bifurcation from $0$, but is somewhat
less satisfactory quantitatively in this case. This branch becomes
unstable around $C=0.08$ due to the collision of one of these eigenvalue
pairs with one of opposite Krein signature, leading once again to a 
quartet of eigenvalues. The details of the subsequent dependence
of this unstable eigenvalue on $C$, both for this branch and for $|1,-e>$, 
depend also on this size of the domain for reasons similar to those 
discussed in \cite{JK99}. The two bottom panels show the unstable 
evolution of the corresponding solutions, illustrating an interesting 
phenomenon particularly in the case of the $|1,-e,e>$ branch. 
The dynamical evolution favors the tunneling of the excitation from
the position of the interface to a nearby site (in this case, mainly
to $n=2$). That is, the interface displaces the solution towards
a position where the environment is more conducive (being surrounded by 
focusing sites) to the existence of a localized pulse solution.

\section{Conclusions}

In the present paper, we have introduced a new setting for the
study of the recent theme of inhomogeneous nonlinear lattices
in nonlinear optics (waveguide arrays) and Bose-Einstein condensates
(optical lattices and superlattices).
The setting consists of an interface between defocusing and focusing
(repulsive interaction and attractive interaction, respectively) 
regions and a transient layer between the two. 

We have focused specifically on the statics and dynamics of coherent waveforms
in the vicinity of this interface, and we have found that their
properties are dramatically modified in comparison with those
expected from the homogeneous case. Some manifestations of these
differences can be quantified in the termination of the principal
pulse branch (for small couplings) or the more prolonged (in parameter
space) persistence of more extended structures in comparison with 
more localized ones. Furthermore, we have shown that the interface
may induce a dynamical tunneling of the structures towards 
locations more favorable for their existence. We have also developed
a systematic methodology based on the adaptation of the considerations
of \cite{DKF05} to inhomogeneous settings and illustrated how to use
these to develop a perturbative treatment of the problem with 
excellent qualitative and good quantitative agreement with the
full numerical results. 

There are many interesting questions that are suggested for
the interface problem we have introduced. 
A prominent one concerns 
the dynamical evolution of localized structures towards the interface
and their interaction (transmission, reflection or trapping) with that region.


\pagebreak
\pagebreak
\pagebreak

\end{document}